\documentclass[prl,twocolumn,showpacs,preprintnumbers,amsmath,amssymb]{revtex4}

\usepackage{graphicx}
\begin{document}

\title{Coexistence of $s$-wave Superconductivity and Antiferromagnetism }

\author{M. Feldbacher$^{1}$, F.F. Assaad$^{1,2}$, F. H\'ebert$^{3}$,
 and G. G. Batrouni$^{4}$}
\affiliation{1. Institut f\"ur Theoretische Physik III, 
   Universit\"at Stuttgart, Pfaffenwaldring 57, D-70550 Stuttgart, Germany,
   Germany. \\2. Max Planck institute for solid state research,
   Heisenbergstr. 1, D-70569, Stuttgart, Germany. \\
   3. Theoretische Physik, Universit\"at des Saarlandes, 66041 
   Saarbr\"ucken, Germany. \\
   4. Institut Non-Lin\'eaire de Nice, Universit\'e de Nice-Sophia
   Antipolis, 1361 route des Lucioles, 06560 Valbonne, France.}

\begin{abstract}
We study the phase diagram of a new model that exhibits a first order
transition between s-wave superconducting and antiferromagnetic
phases.
The model, a generalized Hubbard model augmented with competing 
 spin-spin and pair-pair interactions, was investigated using 
the projector Quantum Monte Carlo method. Upon varying
the Hubbard $U$ from attractive to repulsive we find a first order 
phase transition between superconducting and antiferromagnetic states.
\end{abstract}

\maketitle

Experimental phase diagrams with superconducting  and magnetic phases in 
close proximity  continue to fascinate since the broken symmetry 
states are  incompatible. In particular, a canonical  two-dimensional 
example is the layered organic superconductor 
$\kappa$-(BEDT-TTF)$_{2}$Cu[N(CN)$_{2}$]Cl
which shows a first order
transition between antiferromagnetic (AF) insulating and superconducting (SC) phases
upon varying the pressure.\cite{Lefebvre} Debate continues
about the  nature of the superconducting state. Kanoda et al. argue that 
NMR data are consistent with an unconventional superconducting gap 
with nodal lines.\cite{Kanoda} Conversely, many experimental studies are
rather consistent with s-wave, phonon mediated superconductivity. In particular,
Kini et al. could demonstrate a BCS-like mass isotope effect 
in $\kappa$--S.\cite{Kini} The same material exhibits a strong 
superconductivity induced (acoustic) phonon renormalization. \cite{Pintschovius}
Nodes in the pair order parameter are also incompatible with 
specific heat measurements.\cite{Muller} 

\begin{figure}[t]
\begin{center}
\includegraphics[width=.4\textwidth]{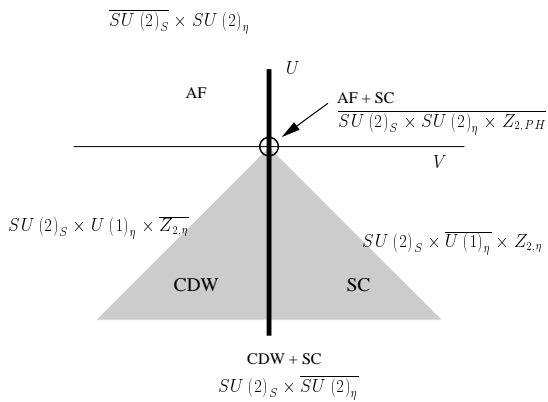}
\end{center}
\caption[]{
The phase diagram for the full Hamiltonian. The shaded region indicates
the parameters accessible to sign-free simulations. For every region 
in the phase diagram the full symmetry of the Hamiltonian is indicated. 
Broken symmetries are marked by bars.}
\label{phase.fig}
\end{figure}

In order to model the scenario of a phase transition between an AF
and an s-wave SC on a half-filled two dimensional
lattice we add to the standard Hubbard model a pair hopping term: 

\begin{eqnarray}
\label{H}
 H & = & -t \sum_{ \left\langle \vec{i}, \vec{j} \right\rangle, \sigma } 
        \left(  c^{\dagger}_{\vec{i},\sigma}  c_{\vec{j},\sigma} + H.c.
        \right) 
           + U \sum_{\vec{i}}n_{\vec{i},\uparrow} n_{\vec{i},\downarrow}\\
    &- & t_p \left( \sum_{\sigma} c^{\dagger}_{\vec{i},\sigma} 
      c_{\vec{j},\sigma} + H.c.  \right)^2 - 4 V \sum_{ \left\langle \vec{i},
        \vec{j} \right\rangle} \eta_{\vec{i}}^z \eta_{\vec{j}}^z  \nonumber
\end{eqnarray}
where $\vec{i}$ labels the sites of a square lattice and the first sum
runs over nearest neighbor bonds. The fermionic spin is given by 
$\vec{S}_{\vec{i}}=\frac{1}{2}\sum_{s,t}c^{\dagger}_{\vec{i},s}\vec{\sigma}_{s,t}c_{\vec{i},t}$
and $\vec{\sigma}$ denotes the Pauli matrices. For a dimer the pair hopping
 term arises exactly when we assume a Su-Schrieffer-Heeger 
electron-phonon interaction 
 \cite{Su80} and integrate out the phonon in the antiadiabatic limit. 
We introduce a particle-hole transformation $\mathcal{P}$  with properties
$\mathcal{P}c_{\vec{i},\uparrow}\mathcal{P} = c_{\vec{i},\uparrow}$, 
$\mathcal{P}c_{\vec{i},\downarrow}\mathcal{P} = 
(-1)^{\vec{i}}c^{\dagger}_{\vec{i},\downarrow}$
and $\mathcal{P}^2=\mathcal{P}$.
Next we define the pairing operators as \cite{Zhang}
$\vec{\eta}_{\vec{i}}=\mathcal{P}\vec{S}_{\vec{i}}\mathcal{P}$, the pair
creation operator $\eta^{\dagger}_{\vec{i}} = 
  (-1)^{\vec{i}}c^{\dagger}_{\vec{i},\uparrow}c^{\dagger}_{\vec{i},\downarrow}$
and $\eta^{z}_{\vec{i}} = (n_{\vec{i},\uparrow}+n_{\vec{i},\downarrow}-1)/2$
 is the total charge. 
The pair hopping term may be recast \cite{Assaad01} as
$  -\left( \sum_{\sigma} c^{\dagger}_{\vec{i},\sigma} 
      c_{\vec{j},\sigma} + H.c.  \right)^2=4 \left(\vec{S}_{\vec{i}}
      \vec{S}_{\vec{j}} + \vec{\eta}_{\vec{i}} \vec{\eta}_{\vec{j}}
      -\frac{1}{4}\right).  $
For the simulations, we have used the 
projector auxiliary field QMC algorithm which is well suited for the
study of ground state properties \cite{Assaad01}. Simulations do 
not suffer from a sign problem in the region of attractive $U < 0$ 
with the additional constraint $U\leq -8|V|$. This is the shaded region 
in Fig.~\ref{phase.fig}. Using particle-hole symmetry the region of 
repulsive $U$ at $V=0$ is also accessible.

Two $SU(2)_{S,\eta}$ symmetry groups are generated by the spin algebra 
$\left\lbrack S^{+}_{\vec{i}},S^{-}_{\vec{j}}\right\rbrack = 
 2 \delta_{\vec{i},\vec{j}} S^{z}_{\vec{i}} $ 
and its particle-hole equivalent for the $\eta$ operators. Note that the 
commutator $\left\lbrack
  \vec{S}_{\vec{i}},\vec{\eta}_{\vec{i}}\right\rbrack=0$
because $S^{\alpha}_{\vec{i}} \eta^{\beta}_{\vec{i}}=0$.
The hopping terms $H_t$ and $H_{t_p}$ are known to be invariant under spin
rotation and also commute with the discrete symmetry $\mathcal{P}$. This 
in turn implies that $\vec{\eta}_{\vec{i}}=
\mathcal{P}\vec{S}_{\vec{i}}\mathcal{P}$ 
also commutes with both hopping terms. 
The symmetries of the Hubbard term become apparent after putting it into 
the form
\begin{equation}
\label{eq:Hubbard} 
       H_{U} = U/3 \sum_{\vec{i}} \left\lbrack 
 \left( \vec{\eta}_{\vec{i}}\right)^2 - \left( \vec{S}_{\vec{i}}\right)^2
 \right\rbrack.
\end{equation}
We can read off the $SU(2)_{S}\otimes SU(2)_{\eta}$ symmetry and observe 
that the Hubbard term changes sign under $\mathcal{P}$ 
transformation. \cite{Zhang}
In Fig.~\ref{phase.fig} the various symmetries for different parameter
regions are displayed.  Bars denote symmetries 
which we numerically find to be spontaneously broken in the thermodynamic
limit. The point $U=V=0$ has the full symmetry: 
$SU(2)_{S}\otimes SU(2)_{\eta}\otimes Z_{2,PH}$ where $Z_{2,PH}$ denotes the 
discrete particle-hole symmetry. 
Here two scenarios are possible: i) Long range order for \emph{both} AF and SC
correlations since one implies the other through $Z_{2,PH}$ symmetry or ii)
The competition between the two broken symmetry states leads to a disordered 
state.
We find that the system realizes the first possibility of coexistence of
s-wave SC and AF. 
Switching
on $U$ breaks the particle-hole symmetry whereas $V$ reduces the 
$SU(2)_{\eta}\rightarrow U(1)\otimes Z_{2,\eta}$.

In the following, we will first discuss the phase transition 
at constant negative $U$ as a function of $V$ and then turn our attention to 
the phase transition at $V=0$ and varying $U$.

In the $U \rightarrow -\infty$ limit the model is equivalent to hard-core 
bosons with repulsive nearest neighbor interactions. At half-filling 
this system experiences a transition from SC to charge-density wave states
(CDW) going across $V=0$. \cite{Troyer} 
The transition point is marked by the higher symmetry $SU(2)_{\eta}$ and 
broken symmetry states may be labeled according to the direction of the 
magnetization. Therefore it is possible to find broken symmetry states
which show a coexistence of CDW and SC order parameters. One can view this 
as a consequence of merging the Ising and the $U(1)$ symmetries into a 
continuous group.

\begin{figure}[t]
\begin{center}
\includegraphics[width=.5\textwidth]{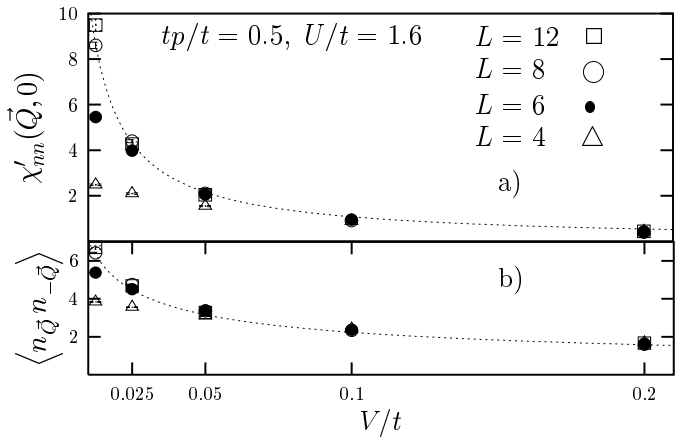}
\end{center}
\caption[]{a) Uniform susceptibility $\chi_{n,n}^{\prime}(\vec{Q},0) 
 \propto \xi^2 \Delta^{-1}$
b) $\left\langle n_{\vec{Q}}\;n_{-\vec{Q}} \right\rangle \propto \xi^2$. The 
dotted line is proportional to results from LSWT.
}
\label{Ueff.DDens.fig}
\end{figure}

\begin{figure}[t]
\begin{center}
\includegraphics[width=.5\textwidth]{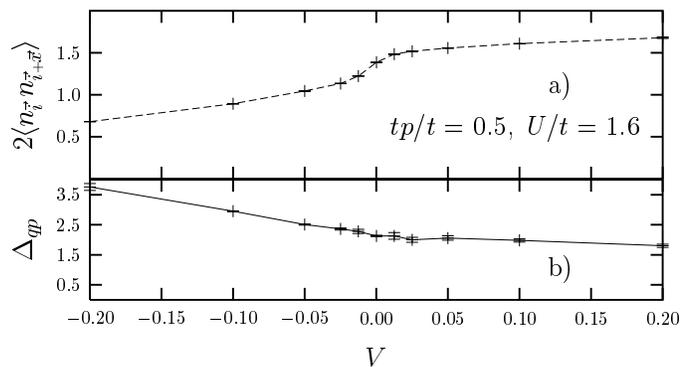} \\
\end{center}
\caption[]{a) Density-density correlations
$2\left\langle n_{\vec{i}} n_{\vec{i}+\vec{x}}\right\rangle = -1/N (\partial F / \partial V) $
where $n_{\vec{i}}=n_{i,\uparrow}+n_{i,\downarrow}$. The derivative of the free energy 
is continuous. b) The quasi-particle gap $\Delta_{qp}$ was obtained by fitting the tail of 
the imaginary time correlation function $G(\tau)$.
}  
\label{DDens_V_Fig.fig}
\end{figure}

\begin{figure}[t]
\begin{center}
\includegraphics[width=.5\textwidth]{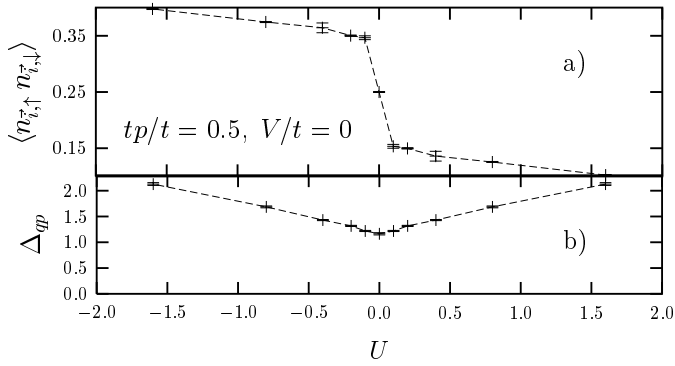} \\
\end{center}
\caption[]{ a) The "Hubbard" correlation 
$\left\langle n_{\uparrow}n_{\downarrow}\right\rangle =1/ N (\partial F/ \partial U)$
has a jump at the transition point $U=0$. The first order transition is
signaled by a discontinuity of the first derivative of the free energy.
b) The quasi-particle gap $\Delta_{qp}$.
}
\label{GApHubb.fig}
\end{figure}

The same behavior is still present for finite attractive $U$. In 
particular our simulations show (see Figs.~\ref{DDens_V_Fig.fig}b and \ref{GApHubb.fig}b) 
that single  particle excitations are
always gapped so we can argue that
the system still renormalizes to the bosonic model.
As we will see, crossing the $V=0$ line at $T=0$, we observe a divergence 
of the correlation length (Fig. \ref{Ueff.DDens.fig})
which points to a second order phase transition. Furthermore, we find no sign
of a discontinuity in the first derivative of the free energy 
$(1/N) \partial F / \partial V = -2 \langle n_{\vec{i}} n_{\vec{i}+\vec{x}}
 \rangle$ (see Fig. \ref{DDens_V_Fig.fig}a).
At the same time we observe a jump for the CDW and SC order parameters at $V=0$.
To understand this transition, we split it up into three parts:
a) $V \rightarrow 0^{-}$, b) $V=0^{-}\rightarrow 0^{+}$
and c) $V \rightarrow 0^{+}$. Part b) is responsible for the jump in the 
order parameters. Let us introduce the two limiting ground states via
\begin{eqnarray}
\label{} 
\Psi_{\mathrm{Ising}} &=& \lim_{V \rightarrow 0^{-}} \lim_{N \rightarrow
  \infty}\Psi(V) \\
\Psi_{\mathrm{XY}} &=& \lim_{V \rightarrow 0^{+}} \lim_{N \rightarrow
  \infty}\Psi(V).
\end{eqnarray}
The ground state in the Ising regime $\Psi_{\mathrm{Ising}}$ has CDW order and 
the SC order parameter is zero. Jumping to positive $V$ and  $\Psi_{\mathrm{XY}}$
the situation reverses. Thus the observed jump in the order parameters
is the simple consequence of forcing the transition from one broken 
symmetry state to the other: 
$\Psi_{\mathrm{Ising}} \rightarrow \Psi_{\mathrm{XY}}$.
Next we look at the critical behavior as we approach the phase transition
from either side. Linear spin wave theory (LSWT) predicts for both transitions
a) and c)  the same square root divergence
for the respective transverse  correlation function. In case c) the equal time
transverse correlation function is 
$\langle n_{\vec{Q}} n_{-\vec{Q}} \rangle$ where 
$ n_{\vec{Q}} = (1/\sqrt{N}) \sum_{\vec{r}} e^{i\vec{Q}\vec{r}} n_{\vec{r}} $, 
$n_{\vec{r}}=n_{\vec{r},\uparrow}+n_{\vec{r},\downarrow}$ and $\vec{Q} = (\pi,\pi)$.
According to LSWT this density-density correlation diverges like  
$\langle n_{\vec{Q}} n_{-\vec{Q}} \rangle = 8 S /\sqrt{V/t_p}$ as 
 $V \rightarrow 0^{+}$. The real part of the uniform susceptibility 
 \begin{equation}
   \chi_{n,n}^{\prime}(\vec{k}=\vec{Q},\omega=0)=\int_{0}^{\infty}d\tau
 \left\langle n_{\vec{Q}} n_{-\vec{Q}}\right\rangle(\tau)
 \end{equation}
picks up the inverse of the gap $\Delta=8\sqrt{2 V t_p}$ and in LSWT 
diverges like  $ \chi_{n,n}^{\prime}(\vec{Q},0)=\sqrt{1/2} S/V$.
Asymptotically the correlation function 
$\left\langle n_{\vec{i}} n_{0} \right\rangle(\tau)$ 
decays exponentially in space and time thus the integrated correlation
functions are proportional to the correlation length and gap
\begin{eqnarray}
  \label{eq:correlationlength}
  \left\langle n_{\vec{Q}} n_{-\vec{Q}} \right\rangle &\propto& \xi^{2}  \\
   \chi_{n,n}^{\prime}(\vec{k}=\vec{Q},\omega=0)      &\propto& \xi^{2} \Delta^{-1}.
\end{eqnarray}
Data for both $ \chi_{n,n}^{\prime}(\vec{Q},0)$ and $\langle
 n_{\vec{Q}}^{z} n_{-\vec{Q}}^{z}\rangle(\tau=0)$ 
are shown in Fig.~\ref{Ueff.DDens.fig} where the dotted line plots the LSWT
 result multiplied with a single constant to account for the reduced moment.

The ordered state in the XY regime has one Goldstone mode and the transverse
 correlations $\langle n_{\vec{Q}} n_{-\vec{Q}} \rangle$ are
 gapped.
Approaching the critical point the gap closes and the 
$\langle n_{\vec{Q}} n_{-\vec{Q}} \rangle$  correlation length
diverges. 
Thus critical behavior reflects the softening of the observed mode which turns into a
Goldstone mode at the point with higher symmetry. Ultimately it is the softening of the
transverse  mode which is responsible for the observed critical behavior.

\begin{figure}[t]
\begin{center}
\includegraphics[width=.5\textwidth]{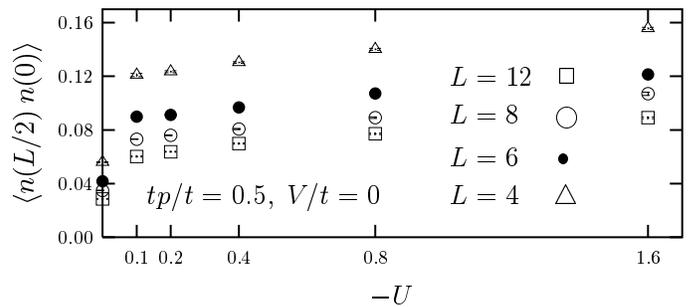} \\
\end{center}
\caption[]{The density-density correlation function  
$ 1/N \left\langle n(L/2) n(0)  \right\rangle  $ at maximum separation
$(L/2,L/2)$.
The correlation function jumps by a factor two once a finite
Hubbard $U$ term breaks the $Z_{2,PH}$ symmetry.}  
\label{Uscan.DDens.fig}
\end{figure}

\begin{figure}[t]
\begin{center}
\includegraphics[width=.5\textwidth]{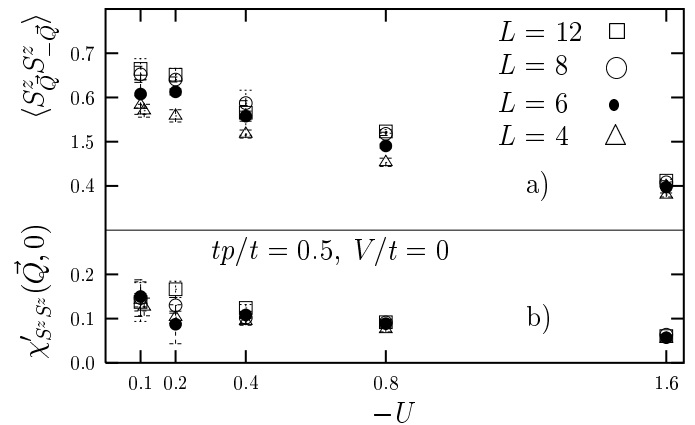} \\
\end{center}
\caption[]{Values for the uniform susceptibility $\chi_{S^z,S^z}^{\prime}( 
\vec{k}=\vec{Q},\omega=0 ) \propto \xi^2 \xi^z$ (lower curve) are well below
the correlation function $\langle S^z_{\vec{k}}\;S^z_{-\vec{k}} \rangle 
(\vec{k}=\vec{Q}) \propto \xi^2$}
\label{Uscan.SSpin.fig}
\end{figure}

Following the $V=0$ line as $U \rightarrow 0$ the bosonic LRO decreases only 
slightly leading to a finite limit at $U=0^+$ which is shown in 
Fig. \ref{Uscan.DDens.fig}. Crossing the $U=0$ point we encounter again
a phase transition which is linked to a point with higher symmetry:
the discrete particle-hole symmetry.
 Using the $Z_{2,PH}$ symmetry, 
one can map this bosonic LRO to an equivalent spin LRO for the repulsive
$U$ case at $V=0$. Precisely at $U=0$ particle hole symmetry implies
that bose correlation functions are equal to their spin counterparts 
\begin{equation}
\label{eq:spinispair}
\left\langle S^{\alpha}_{\vec{k}+\vec{Q}}S^{\alpha}_{-\vec{k}-\vec{Q}}
\right\rangle =
\left\langle \eta^{\alpha}_{\vec{k}}\eta^{\alpha}_{-\vec{k}}
\right\rangle.
\end{equation}
This explains a jump by a factor two for the density-density correlations as
one moves from the symmetric point to a finite $U$ which is visible in 
Fig. ~\ref{Uscan.DDens.fig}.
The emergence of a Goldstone mode above was associated with the critical behavior
of the phase transition. As we are now approaching the $U=0$ point the only 
continuous symmetry available which is not associated with the bosonic LRO is
the spin symmetry. In Fig.~\ref{Uscan.SSpin.fig} we compute the
correlation length for $\langle S^{z}_{\vec{Q}} S^{z}_{-\vec{Q}} \rangle$. 
This quantity shows no sign of divergence at the phase transition.
Contrary to what happened in the SC to CDW transition we may classify this 
transition as a first order level crossing transition. From 
Fig.~\ref{GApHubb.fig} we equally see that the derivative of the free energy with respect 
to $U$ is discontinuous. The free energy as a function of double occupancy has
a plateau which marks a region of phase coexistence.

\begin{figure}[t]
\begin{center}
\includegraphics[width=.5\textwidth]{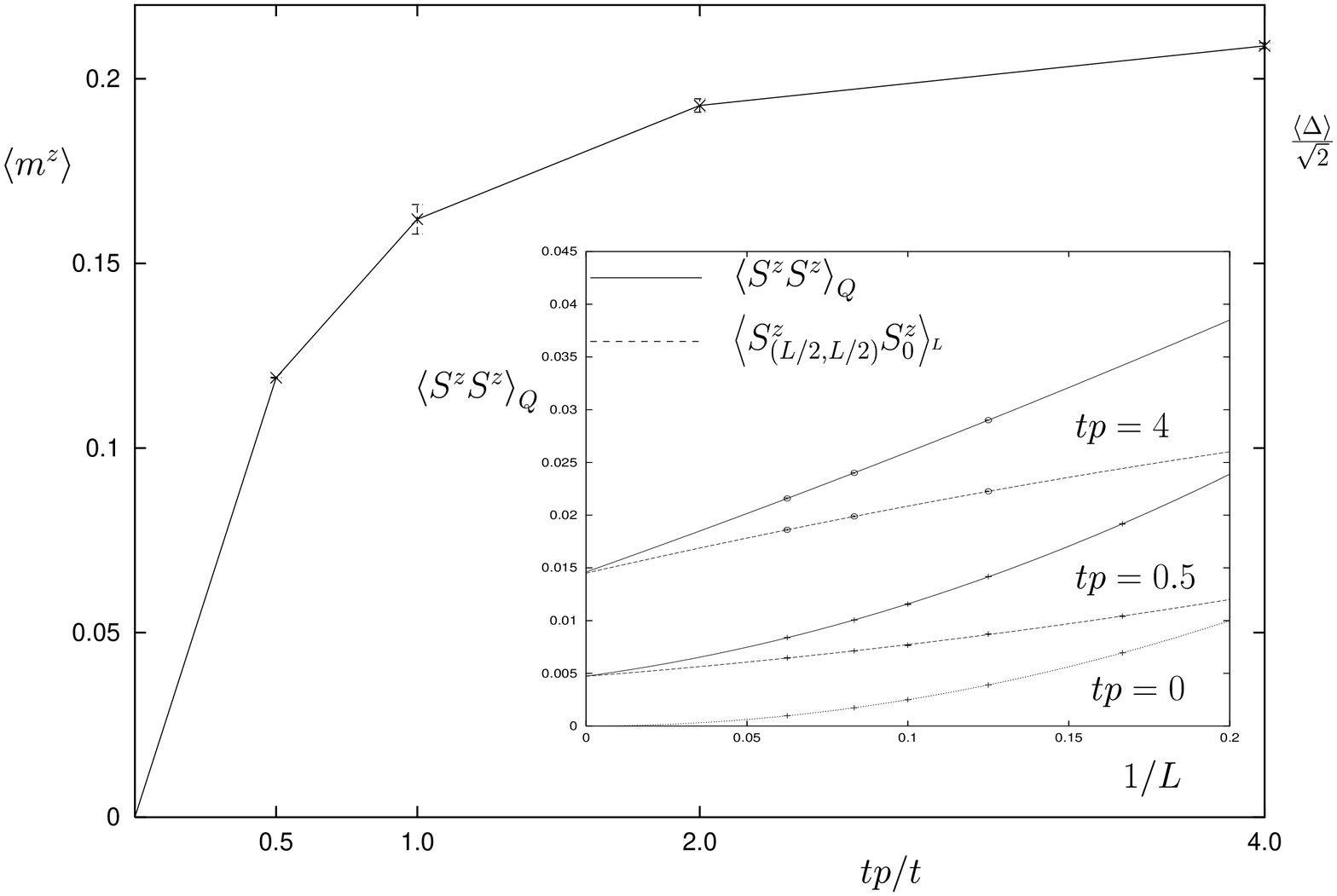} \\
\end{center}
\caption[]{The staggered moment $ m_s $ which coexists with 
the superconducting order parameter $ \Delta = m_s \sqrt{2}$ plotted as a function of
$t_p\;(U/t=V/t=0)$. The inset shows finite size extrapolation for $t_p=4$ and $t_p=0.5$. 
}
\label{Mag.fig}
\end{figure} 

For the point $U=V=0$ we find coexistence of the spin and pair order parameters.
After finite size extrapolation the results of Fig.~\ref{Mag.fig} are consistent with 
a finite value of the magnetization $m_s$ up to weak  coupling ($t_p/t << 1$).
At the same time the superconducting order parameter $\Delta$ is also nonzero 
and the measured value $\Delta = \sqrt{2} m_s $ agrees with Eq.~\ref{eq:spinispair}. 
The observed coexistence of AF and SC order is due to the  
special particle-hole symmetry $Z_{2,PH}$ at the point $U=0$  which renders 
the energies of  both the AF and SC states degenerate.

\begin{figure}[h]
\begin{center}
\includegraphics[width=.5\textwidth]{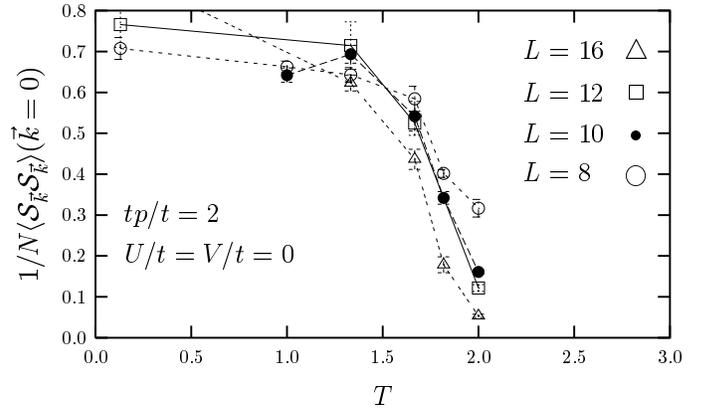} \\
\end{center}
\caption[]{The $1/N \left\langle \mathcal{S}_{\vec{k}}\mathcal{S}_{\vec{k}}\right\rangle
(\vec{k}=0)$ correlations show that at least up to $T\leq 1.8 t$ particle-hole
symmetry is broken.
}
\label{Ising.fig}
\end{figure}

In order to understand how the 
$SU(2)_{S}\otimes SU(2)_{\eta}$ symmetries are spontaneously broken we also
need to discuss whether the discrete particle-hole symmetry is broken and how 
this affects the spontaneous breaking of the continuous symmetries.
We introduce the following pseudo-spin order parameter
\begin{eqnarray}
\label{eq:isingorderpar}
\mathcal{S}_{\vec{i}} & = & \left( (2 S_{\vec{i}}^z )^2 -
(2 \eta_{\vec{i}}^z)^2 \right) \left\{
\begin{array}{c}
+1: \left| \uparrow\right\rangle , \left| \downarrow\right\rangle \\
 -1: \left| \uparrow\downarrow\right\rangle , \left| 0 \right\rangle
\end{array}
\right. ,\\
\label{isingzero}
\left\langle \mathcal{S} \right\rangle &=& 
\left\langle \mathcal{PSP}\right\rangle =
-\left\langle \mathcal{S} \right\rangle
\end{eqnarray}
which has to be zero as long as the $Z_{2,PH}$ symmetry is not broken as 
assumed in Eq. (\ref{isingzero}). A positive (negative) value for 
$\left\langle\mathcal{S}\right\rangle$ indicates a majority of spins (pairs)
in the system. Plotting the uniform correlations 
$1/N \left\langle \mathcal{S}_{\vec{k}}\mathcal{S}_{\vec{k}}\right\rangle(\vec{k}=0)$ as a function of 
temperature, one sees in Fig.~\ref{Ising.fig} that below
a temperature $T_c \sim 1.8 t$  particle-hole symmetry is indeed broken.
Hence below $T_c$ and  in the thermodynamic limit  the $Z_{2,PH}$  particle-hole
symmetry is broken  and depending on the orientation of the pseudo-spin,
$\mathcal{S}_{\vec{i}}$, the SC or AF ground state will be chosen.

In conclusion we have considered a model with competing interactions. 
The main ingredients in the model are on the one hand 
the Heisenberg and Hubbard terms which lead to a localization of single electrons
thereby producing antiferromagnetic insulating states  and on the other hand, 
pair hopping and nearest neighbor attractive terms which lead to on-site s-wave 
superconducting states.  Based on the symmetries of the model as well as quantum Monte 
Carlo simulations we have shown that the model has a variety of phase transitions. 
In particular we have found a first order phase transition between an s-wave
superconducting state and antiferromagnetic Mott insulator at a high symmetry point
($U=V=0$) where the two phases coexist.

We wish to thank the HLR-Stuttgart for generous allocation of computer
time, the DFG for financial support (grant numbers AS 120/1-3 , AS
120/1-1 ) as well as a joint Franco-German cooperative grant
(PROCOPE).

\end{document}